# Enhanced Non-Ohmic Drain Resistance of 2DFETs at Cryogenic Temperature

Kwok-Ho WONG[1#], Mansun CHAN[1†]

[1]Department of Electronic and Computer Engineering, The Hong Kong University of Science and Technology, Hong Kong, China

KEYWORDS: Two-dimensional materials, Schottky barrier FET, cryogenic temperature

ABSTRACT

The contact issue for two-dimensional (2D) materials-based field-effect transistors (FETs) has drawn enormous attention in recent years. Although ohmic behavior is achieved at room temperature, the drain current of 2DFETs shifts from ohmic to non-ohmic behavior at cryogenic temperatures. In this work, we demonstrate that the shift is attributed to the asymmetric current reduction at the metal-semiconductor contact at low temperature. Under low drain bias, carriers tunnel from the source to the channel but diffuse to the drain side due to the channel-to-drain barrier, resulting in the current suppression. By studying the property of ohmic metal-semiconductor contact at different temperatures, we analyzed the mechanisms behind this phenomenon and the dependence on metal-to-semiconductor barrier height. The work opens the semiconductor physics of 2D material contact at cryogenic temperature and the importance of contact metal selection in the development of 2DFET at cryogenic temperature.

INTRODUCTION

Two-dimensional materials have emerged as promising candidates to address the scaling limitations of conventional transistor technology.[1-4] A critical factor in the success of 2D transistors is the achievement of low contact resistance. Significant research efforts have been dedicated to reducing the contact resistance in these devices.[5-8] Unlike traditional silicon-based transistors, the distinct interface between the metal and the 2D semiconducting material often results in the formation of Schottky contacts.

While the contact resistance at the source is generally regarded as having a more pronounced impact on overall device characteristics,[9-12] the influence of contact resistance at the drain is frequently overlooked. However, some reported data indicate that high contact resistance at the drain can affect the overall drain current versus drain voltage relationship, even when source resistance, as determined from transfer characteristics, is minimized.[13-15] This effect becomes particularly pronounced at low temperatures, where drain resistance can dominate in cryogenic conditions.

In this work, we aim to provide a physical explanation for the enhanced non-linear drain resistance observed at cryogenic temperatures and demonstrate that its effect can more significantly affect the current voltage characteristics than the source metal-to-semiconductor contact.

ENHANCED DRAIN RESISTANCE

This study utilizes a generic 2D transistor structure, depicted in Figure 1a, featuring a bottom gate configuration to align with most existing data. This structure is sufficiently versatile to accommodate similar configurations. As illustrated, the transistor comprises three distinct components: a Schottky diode at the source, the main field effect transistor, and a Schottky diode at the drain, as detailed in Figure 1b.

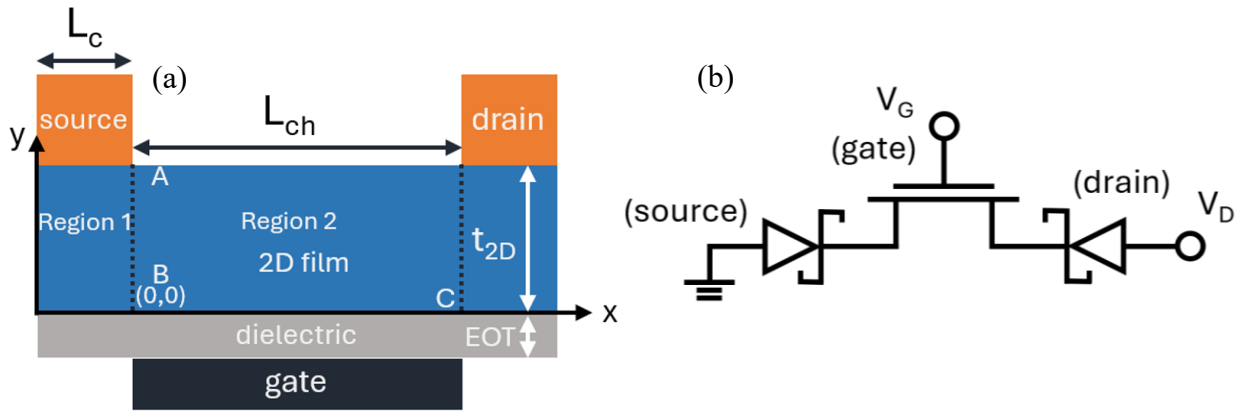

Figure 1. (a) Physical structure of 2DFET. (b) 2DFET is considered as SBFET and can be represented schematically by two Schottky diodes (Metal-semiconductor contact) in series with a junctionless transistor.

To experimentally validate the enhanced drain resistance effect, 2D $MoS_2$ field effect transistors (FETs) were fabricated with the cross-section presented in Figure 1a. A high-k metal gate (HKMG) is implemented, with a Ti (3 nm) / Au (20 nm) local bottom gate (LBG) and around 30 nm ALD $Al_2O_3$ bottom gate dielectric. To achieve linear output characteristics at room temperature, Ni is strategically selected as the contact metal, with a reported contact resistance of 500Ω and a Schottky barrier height of 100meV at 300K.[16-17]

The measured transfer and output characteristics are shown in Figure 2a and Figure 2b, respectively. The transfer curves exhibit typical behaviors at all temperatures and the output curve at room temperature shows normal transistor behavior, confirming the proper functionality of the $MoS_2$ FET. However, the output characteristics experience significant degradation when the temperature is reduced, exhibiting remarkably higher resistance at low drain voltages and a delayed turn on. The observed $I_D$-$V_D$ curves indicate that the on-region drain current is primarily constrained by the drain resistance at low temperatures rather than the source resistance.

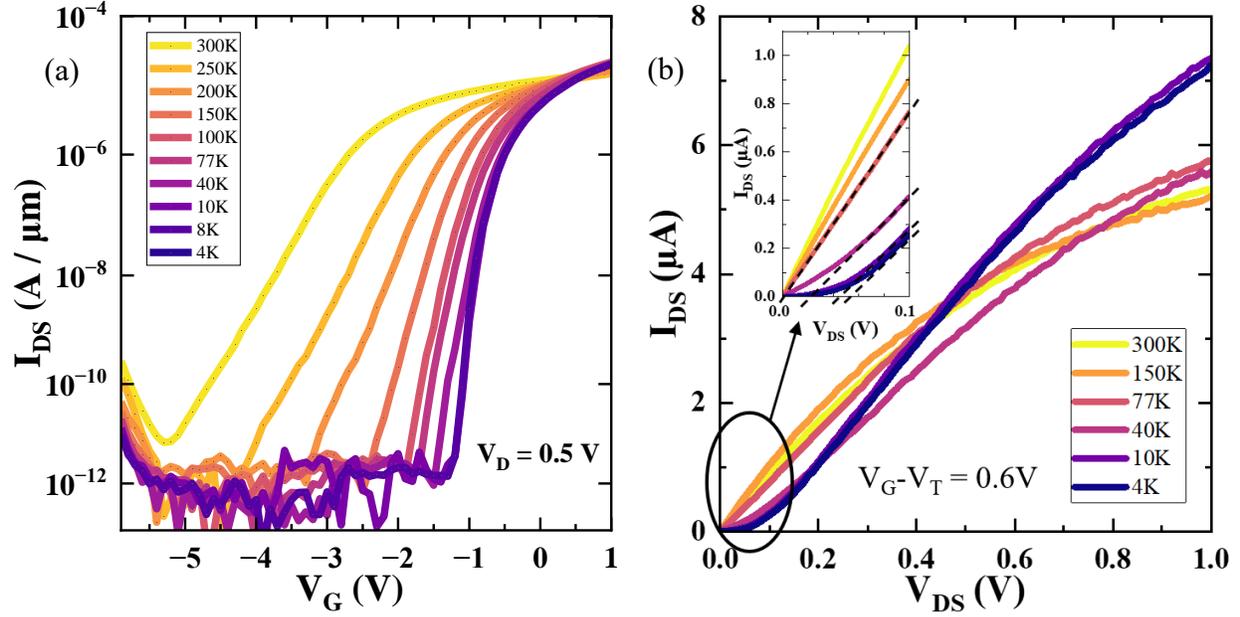

Figure 2. (a) Transfer and (b) output characteristics of the fabricated $MoS_2$ FET with temperatures ranging from 300K to 4K

ANALYSIS OF ENHANCED DRAIN RESISTANCE

To study the physical mechanism behind the enhanced drain resistance, we investigated the temperature effect on a metal-semiconductor (MS) contact. The contact interface between the metal and a thin n-type semiconductor film with $MoS_2$ material properties is analyzed at various temperatures using Silvaco TCAD Simulation. In order to form an ohmic contact and match the experimental condition, the metal-to-semiconductor barrier of the contact is selected to be 100meV. [16-17]

The simulated current of the MS contact on a linear scale at different temperatures is shown in Figure 3a. At room temperature, the simulated MS contact displays a symmetrical ohmic behavior. When the temperature is reduced, the reverse current with electrons moving from the metal to the semiconductor remains the same. However, the forward current with the electrons moving from the semiconductor to the metal is reduced together with the reduction of temperature.

To explain the observation, the conduction band energy (CBE) of the MS contact at different temperatures under a reverse bias of -0.1V and forward bias of 0.1V is plotted in Figure 3b and Figure 3c. At low temperatures, the barrier height from the metal to the semiconductor increases slightly[18] and the conduction band of the semiconductor moves closer to the Fermi level[19] when a bias is applied. Figure 3b also illustrates the carrier conduction under reverse bias, which is determined by tunneling and remains unaffected by temperature changes. In addition, the increase in barrier height is offset by the narrowing of the tunneling barrier which cancels the effect of each other. Consequently, the reverse current remains unaffected by temperature.

In forward bias, the current at room temperature or high bias voltage is driven mainly by drift due to the E-field at the junction forcing the electrons to move from the semiconductor to the metal side. However, as shown in Figure 3c, at low temperatures, the band-bending direction changes because the conduction band shifts towards the Fermi level. Therefore, the current conduction becomes diffusion to overcome the built-in potential at the contact at low bias until the applied voltage is sufficiently larger to reverse the direction of the band bending. Once the drift current dominates, the forward current at low temperatures increases with the reduction of temperature due to the increases in carrier mobility.[20] The switch in the current conduction mechanism can be more clearly seen when we plot the low bias region of I-V characteristics in Figure 3a in log scale, which is shown in Figure 3d. We clearly observe the two different conduction mechanisms of which diffusion dominated a low bias until the drift current picks up at a higher bias voltage. The diffusion current decreases with temperature due to the higher potential to overcome as well as low thermal energy so that the number of electrons able to overcome the potential barrier also reduces.

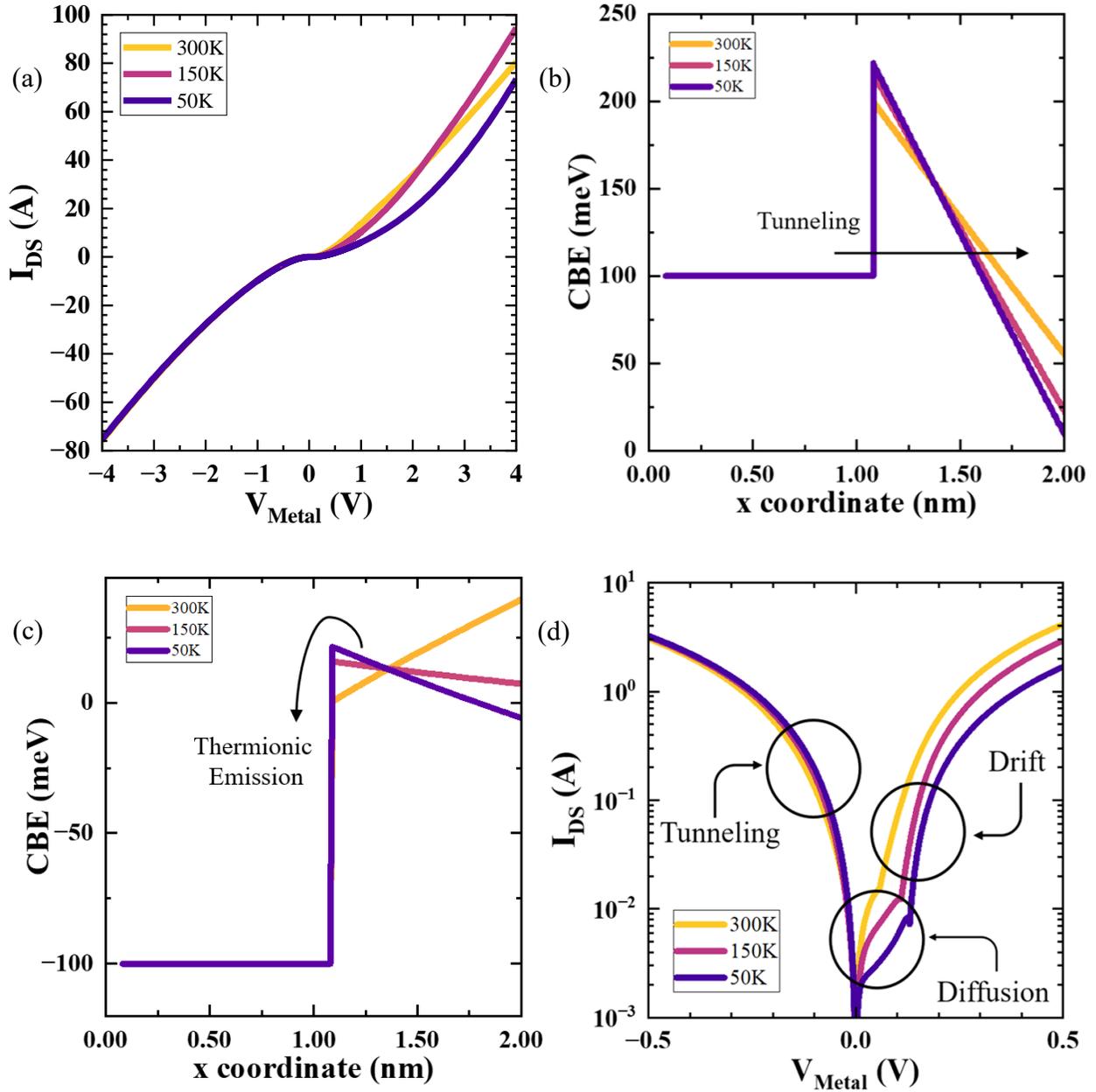

Figure 3. (a) $I_D$-$V_D$ characteristics of an ohmic MS contact with temperatures ranging from 300K to 50K; The conduction band energy (CBE) in an ohmic MS contact with different temperatures (b) in reverse bias condition ($V_{Metal}$ = -0.1V) and (c) in forward bias condition ($V_{Metal}$ = 0.1V). (d) log($I_D$)-$V_D$ characteristics of an ohmic metal-semiconductor contact with temperatures ranging from 300K to 50K.

Following the examination of the temperature effect on the MS interface, we simulated a MoS$_2$-based transistor with the physical structure shown in Figure 1a. To explain the impact on carrier motion at different temperatures, the conduction band of a MoS$_2$ FET is plotted in Fig 4(a). In this figure, the gate voltage is adjusted to give the same ($V_G$-$V_T$) for all the curves to eliminate the effect of temperature to the current conductor as a result of the threshold voltage shift. It is observed that the source-to-body condition is not affected by the change in temperature while the barrier for the electron to move from the channel to the drain increased with the reduction of temperature. Therefore, the carrier conduction is not limited by the source-to-body carrier injection, but the channel-to-drain conduction.

To overcome the increased channel-to-drain potential barrier, a higher drain voltage has to be applied. The simulated effect of drain voltage on the channel to drain barrier height is shown in Figure 4b. The channel-to-drain potential barrier, $\varphi_{db}$, is defined as the potential difference between the conduction band edge at the metal/2D interface and at the 2D/dielectric interface of the drain, labeled in Figure 4b. When a positive $V_{GS}$ ($V_{GS} > V_T$) is applied, the MoS$_2$ FET turns on and the conduction band of the 2D semiconductor bends downward, indicating an accumulation of electrons in the 2D/dielectric interface. As such, two identical Schottky diodes are connected back-to-back when $V_{DS} = 0$.

When $V_{DS}$ is small, for example less than 50mV, the applied voltage is insufficient to eliminate the drain barrier ($\varphi_{db} > 0$), resulting in blocking electron flow from the channel to the drain side and the suppression of the drain current. Although electrons are injected by tunneling from the source side to the body, only a small number of carriers diffuse to the drain side due to the presence of the barrier and the lack of sufficient thermal energy to overcome it.

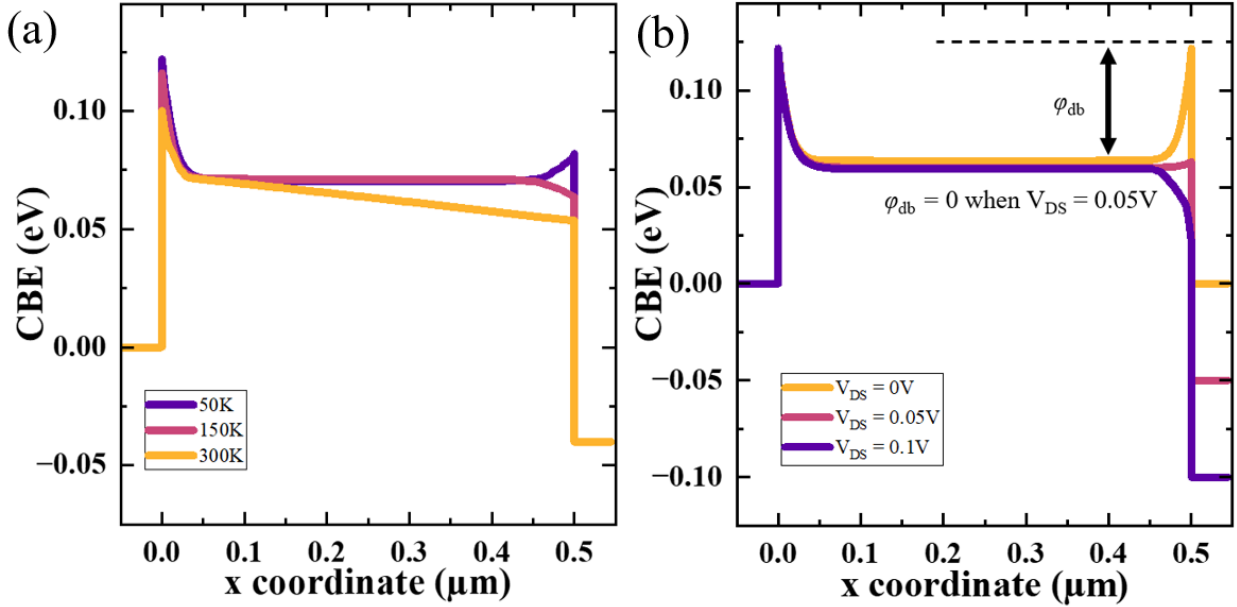

Figure 4. (a) The conduction band distributions in 2DFET with different temperatures at $V_{DS}$ = 40mV; (b) The conduction band distributions in 2DFET at 50K with increasing drain bias.

When $V_{DS}$ is further increased, the barrier ($\varphi_{db} < 0$) is eliminated, allowing electrons to flow to the drain without obstruction. The conduction mechanism changes from diffusion to drift, aligning with the discussion of the MS contact under the forward bias condition. Eventually, as in standard MOSFETs, an increase in $V_{DS}$ leads to the linear and saturation regions.

To quantify the effect of temperature on the drain contact induced non-linearity, the drain voltage to cause the band bending at the channel region to change direction is an important measure. Based on the new understanding of the carrier conduction mechanism at the source and drain junction, we simulated the I-V characteristics of a $MoS_2$ FET and the I-V characteristics are shown in Figure 5a. Herein, we assumed that the 2DFET is fully turned on so that the channel resistance is relatively small. Under this assumption, the channel behaves as a resistor before the carrier enters the velocity saturation and with its resistance decreasing as the temperature is lowered due to the increase in mobility. As shown in Figure 5a, the simulated characteristics show a linear to non-linear transition, aligning well with experimental results.

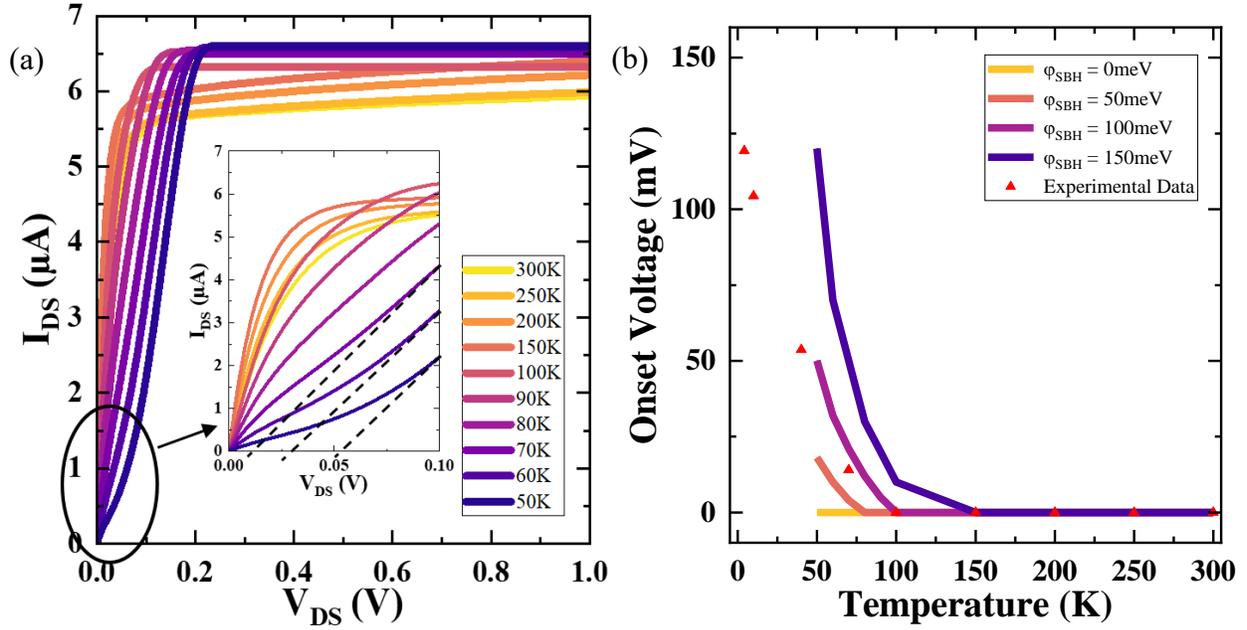

Figure 5. (a) $I_D$-$V_D$ characteristics of a simulated 2DFET (100meV barrier height) with temperatures ranging from 300K to 50K; (b) Temperature dependence of a simulated 2DFET onset voltage.

Moreover, by linear extrapolation, the onset of the drain junction turn-on is shown in Figure 5b indicating a good match between the simulation result and experimental data. Notably, the turn-on voltage can be effectively reduced by minimizing the Schottky barrier height.

CONCLUSION

In conclusion, we have reported the enhanced drain resistance of the 2DFET at cryogenic temperature due to the channel-to-drain barrier. The analysis of the temperature effect on MS contact indicates that the barrier limits the carrier flow, leading to the current suppression. Under the low drain bias, the results of the simulated transistor confirm that the carriers tunnel from the source to the body, and diffuse to the drain region because of the barrier. Therefore, minimizing the Schottky barrier height is essential for preserving linear output characteristics during cryogenic operation. The insights gained from this research not only advance our understanding of semiconductor physics in 2D materials but also provide valuable insights for advancing cryo-CMOS technology.

---

AUTHOR INFORMATION

**Corresponding Author**

# Email: khwongby@connect.ust.hk,

† Email: mchan@ust.hk

**Author Contributions:**

Kwok-Ho Wong confirms sole responsibility for the following: study conception and design, data collection, analysis and interpretation of results, and manuscript preparation. Mansun Chan reviewed the results and approved the final version of the manuscript.

**Notes**

The authors declare no competing financial interests.

**ACKNOWLEDGMENT**

This work was supported by the General Research Fund (GRF) 16203222 from the Research Grants Council (RGC) of Hong Kong.